\documentclass[10pt]{article}
\usepackage{amsfonts}
\usepackage{amsmath}
\usepackage{hyperref}
\usepackage{wick}
\usepackage{epsfig}
\def\arXiv#1{\href{http://xxx.lanl.gov/abs/#1}{arXiv:#1}}
\def\Dir{\,\,{\raise.15ex\hbox{/}\mkern-12mu D}}

\newcommand{\bb}{\begin{equation}}
\newcommand{\bqn}{\begin{eqnarray}}
\newcommand{\eqn}{\end{eqnarray}}
\def\be{\begin{equation}}
\def\ee{\end{equation}}
\def\beq{\begin{eqnarray}}
\def\eeq{\end{eqnarray}}
\newcommand{\pp}{\partial}

\def\d{\delta}

\def\a{\alpha}
\edef\b{\beta}

\def\m{\mu}
\def\n{\nu}

\def\({\left(}
\def\){\right)}
\def\[{\left[}
\def\]{\right]}
\def\mI{\mathbb I}
\def\I{P}

\def\sqN{\sqrt{(N-25)(N-1)}}

\begin{document}

\title{
Families of Commuting Integrals of Motion with Certain Symmetries\\
\ \\
}
\author{
Evgueniy S. Vitchev
\footnote{vitchev@physics.rutgers.edu}
\ \\
\\
{\it Department of Physics and Astronomy, Rutgers University}\\
{\it Piscataway, NJ 08855-0849, USA}
}
\date{}
\maketitle
\centerline{\bf Abstract}
{\center{Some results for commuting local integrals of motion for free
	bosonic theories are
  presented, which can be applied in construction of integrable boundary
  states. Usable program code for calculation of these integrals of
  motion is also introduced.  }}

\section{Introduction}
In the set of theories in two 
dimensions, conformal ones are fixed points \cite{Polyakov:1970xd} with respect to
Renormalization Group (RG)
transformations \cite{Wilson:1973jj}. Perturbing of such a theory with certain
relevant operator yields integrable flow, i.e. a trajectory of ``deformed''
theories,
which leads to another fixed
point,
 which is usually called ``infrared fixed point''. The
intermediate theories on such a trajectory are integrable, meaning that
a countable subset of the integrals of motion (IM) from the original theory
(which is sometimes called ``ultraviolet fixed point'') survive the
relevant perturbation.  In the last several decades there has been a substantial progress in the
understanding of the relation between Conformal Field Theory (CFT) and
Integrable Quantum Field Theory (IQFT) in two dimensions \cite{Bazhanov:1994ft}
\cite{Bazhanov:1996dr} \cite{Bazhanov:1998dq}
\cite{Zamolodchikov:1989zs} \cite{Zamolodchikov:ti} \cite{Zamolodchikov:gt}. 

A conformal field theory is characterized by its chiral algebra
$\cal A$. The algebra $\cal A$ can be either the Virasoro algebra
 or an
extended algebra which includes the Virasoro algebra as a
subalgebra. The Virasoro 
algebra is generated by the holomorphic 
component 
$T$ 
of the energy-momentum tensor. The antiholomorphic component of the
energy-momentum tensor $\bar T$ correspondingly generates another
(antiholomorphic) copy of the Virasoro algebra.
From now on we will
restrict our attention to the holomorphic part of the theory, assuming
that the theory possesses reflection symmetry and therefore all the
reasoning about the holomorphic sector pertains to the antiholomorphic
one as well.

One of the fundamental characteristics of an integrable field theory is the
presence of a countable set of commuting integrals of motion $\{\mathbb
I_s\}$,
\be
[\mathbb I_s,\mathbb I_{s'}]=0,
\ee
where $s$ is the spin of the particular integral. The IM can be given as
integrals of the corresponding densities (currents):
\be\label{intdens}
\mathbb I_s=\oint[P_{s+1}+\Theta_{s-1}d\bar z].
\ee
Note that for a conformal theory the second term in \eqref{intdens} is zero.

An important direction of development is the study of CFT and IQFT with
boundary \cite{Cardy:ir} \cite{Cardy:bb} \cite{Cardy:gw}. One of the
motivations has been the 
application to string 
theory and $D$-branes \cite{Moore:2003vf} \cite{Witten:1991yr}
\cite{Shatashvili:1993ps}. The effect of a particular boundary condition can
be 
given by a boundary state $|B\rangle$, which is a vector in the space of
states of the bulk theory
$${\cal H}=\int\limits_j{\cal F}_j\otimes\bar{\cal F}_j,$$
 where
${\cal F}_j $ and ${\cal \bar F}_j$ are the representations of the left
and right copy of the chiral algebra $\cal A$ labelled by $j$.

Even if a bulk theory is conformal, it is
possible to have a non-conformal, but still integrable boundary
condition, which permits the survival of infinitely many integrals of
motion \cite{Ghoshal:tm}. Here by ``survival'' we mean that the boundary
condition 
preserves the quantity $\mathbb I_s-\bar {\mathbb I}_s$ in the sense
that classically there is no flow of this integral of motion across the
edge of the worldsheet. In quantum version that means
\be\label{bcompat}
(\mathbb I_s-\bar{\mathbb I}_s)|B\rangle=0.
\ee
This means that any boundary state can be given as a linear combination
of eigenstates $\vert\a,j\rangle\in{\cal F}_j$ of the integrals of motion :
\be
|B\rangle=\int\limits_{j}\sum_\a
 B_\a^j(\kappa)|\a,j\rangle\otimes\overline{|\a,j\rangle}, 
\ee
where the coefficients in fact depend on the energy scale $\kappa$, and
$\a$ labels the eigenstates.

It is possible to identify between states in ${\cal F}_j\otimes\bar
 {\cal F}_j$
and operators in the space of linear maps ${\cal F}_j\to \bar {\cal
 F}_j$,
therefore we can associate a boundary operator $\mathbb B$ to the
boundary state $|B\rangle$:
\be
{\mathbb B}(\kappa)=\int\limits_{j}\sum_\a
 B_\a^j|\a,j\rangle\overline{\langle\a,j|}. 
\ee
In some cases the boundary operator can be expanded like
\be
\log {\mathbb B}(\kappa)=\log{\mathbb B}_0+\sum_s
  \frac{{\mathbb I}_s}{\kappa^s}.
\ee
Thus, if we know a particular family of commuting charges, and
find a boundary condition compatible with \eqref{bcompat}, then we can
describe the associated boundary state. Also, it can be inferred that the
partition function for such a theory, which equals to the overlap of the
boundary state with the radial quantization vacuum,
\be
Z(\kappa)=\langle B(\kappa)|0\rangle,
\ee
can be similarly expanded in powers of $\kappa$, whith coefficients
proportional to the vacuum expectation values of the commuting IM:
\be
\log Z=\log Z_0+\sum_s\frac{I_s}{\kappa^s},
\ee
where $I_s=\langle\mathbb I_s\rangle$ are the vacuum expectation values of the
 IM and $Z_0$ is the partition
 function of the infrared fixed point.

Another piece in this mosaic was noticed originally by Dorey and Tateo
\cite{Dorey:1998pt} and corroborated later in other works
\cite{Bazhanov:1994ft} \cite{Bazhanov:1996dr} \cite{Bazhanov:1998dq},
namely, that there exists a peculiar relation between integrabe field
theories and the spectral theory of certain linear ordinary differential
equations (ODE). This relationship, sometimes called the
``ODE/IM correspondence'', is still not very clear, but nevertheless
there is substantial evidence for it. In particular, for many integrable
field theories it has been shown the existence of a special ODE, whose
solutions' Wronskian reproduces with amazing exactness the functional
dependence of overlapping amplitudes of the IQFT on the theory's
parameters.

In this context, the problem of study and classification of families of
commuting integrals of motion $\{\mathbb I_s\}$ is not uninteresting. As
a first step in that direction one can try to find such families for the
case of free bosonic theory in the bulk. If such ones are found (rather 
``when'' than ``if'', e.g. we know of the existence of the KdV-Virasoro
family of IM, generated by $T$ and its composite fields, and others are
expected), we can try to 
identify what possible boundary condition is compatible with that
family, and find out can an ODE be written in the context of the ODE/IM
correspondence. Although this task by itself is not so small (unless/until by
some deeply nontrivial means we gain insight into the problem of general
classification of families of IM for free bosonic theory), plausible
future steps could be the study supersymmetric generalizations or
massive backgrounds.

The goal of the present work is to present a method for calculating of
commuting integrals of motion for free bosonic theory. The only
limitation for this method is the 
problem of solving large over-defined square(bilinear) homogeneous
algebraic sytems. We present also
some results\footnote{It is well-known that free theories have infinite
  series of commuting integrals of motion which are quadratic in the
  fields, but the aim of this work is the study of other, less trivial
  families.} 
which were obtained or verified by applying this
method. For $N$ 
bosonic fields with $O(N)$ symmetry, this includes the (already known)
KdV-Virasoro series \cite{Bazhanov:1994ft} \cite{Sasaki:1987mm}
\cite{Eguchi:1989hs}, the ``$O(N)$'' series 
recently found in 
\cite{Lukyanov:2003rt}, and another series,
which is related to quantum Boussinesq theory
\cite{Bazhanov:2001xm}. 
For the case of two fields having $\mathbb
Z_2\times\mathbb Z_2$ symmetry, we obtained a deformation of the $O(N)$
case, which was identified as related to the paperclip boundary
condition in a previous work \cite{Lukyanov:2003nj}. In addition, for
the case of $N+1$ fields with symmetry $\mathbb Z_2\times O(N)$,
obtained were three new families, which in some sense generalize
particular cases of the paperclip family.
We will also
introduce the software 
implementation 
of the method in question which was developed in the course of this
reserch, which could be easily adapted to cases of other symmetries as
well as other problems involving computation of operator product expansions.

\section{Calculating the commuting local IM}\label{mathbackground}
For our purposes it is convenient to consider the theory on a
cylinder with complex coordinates $v,\bar v=x\pm iy$, where $x\equiv
x+2\pi R$,  
$y\in[0,\infty)$. In this case the bosonic integrals of motion will be
homogeneous.
Let us consider the 
space
 $\cal B$ of bosonic currents generated by linear combinations and
 normal ordered products (at the same point $v$) of
\be
\pp^nX^j(v),\,n\in\mathbb N,\,j\in\{1,\ldots\,N\}.
\ee
Here and further, for simplicity,  we use the notation $\pp\equiv\pp_v$. 
For any current of spin $s$ $P_{s}\in{\cal B}$ we can construct the
corresponding charge of spin $s-1$:
\be
\mathbb I_{s-1}=\oint\limits_C dv\, P_{s},
\ee
where the contour $C$ is taken around the cylinder, and the spin $s$
equals to the number of derivatives in $P_s$.
\begin{figure}[h]
\begin{center}
\epsfxsize=0.2\textwidth
\epsfbox{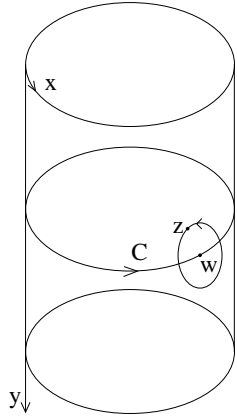}
\end{center}
\caption{Semi-infinite cylinder}
\label{lfig1}
\end{figure}
We denote the algebra of the charges ${\cal Q}$, which is a sum of
spin-$s$ subspaces
\be
  {\cal Q}=\bigoplus_{s=1}^{\infty} {\cal Q}^{s-1}.
\ee
Likewise, ${\cal B}$ is a sum of subspaces (not subalgebras!):
\be
  {\cal B}\cong\bigoplus_{s=1}^\infty{\cal B}^{s}
\ee
In fact, the algebra of charges is isomorphic to the space
 of currents
modulo total derivatives:
\be
{\cal Q}\cong{\cal B}/\pp.
\ee
The same can be stated about the subspaces:
\be
{\cal Q}^{s-1}={\cal B}^{s}/\pp{\cal B}^{s-1}.
\ee
The commutator of two charges,
$\mI_{s_1-1}\in{\cal Q}^{s_1-1}$ and
$\mI_{s_2-1}\in{\cal Q}^{s_2-1}$, is
\be
[\mI_{s_1-1},\mI_{s_2-1}]=
 \oint\limits_C dw\oint\limits_wdz\,P_{s_1}(z)P_{s_2}(w).
\ee
It is easy to check that
\beq
   [\mI_{s_1-1},\mI_{s_2-1}]&=&2\pi
   i\oint\limits_C dw\,J_{s_1+s_2-1}(w)\nonumber\\
&=&2\pi i R_{s_1+s_2-2},
\eeq
where $J_{s_1+s_2-1}(w)\in{\cal B}^{s_1+s_2-1}$ is the residue of the
operator product expansion (OPE) of $P_{s_1}\in{\cal B}^{s_1}$ and
$P_{s_2}\in{\cal B}^{s_2}$, and $R_{s_1+s_2-2}\in{\cal Q}^{s_1+s_2-2}$
is the corresponding
``charge''. In this paper I use a not very standard normalization of the
fields, so the Wick pairing is given by
\be
\overwick{{\pp} {X}^i(z)\pp {X}{^j(w)}}{\contract{\pp}{X}{X}{^j(w)}}
 =\frac{\d^{ij}}{(z-w)^2}.
\ee
To go to a more physical normalization, like the one in \cite{Lukyanov:2003nj},
one needs to do the substitution $X^j\to i\sqrt2X$. Likewise, the
momenta in the vacuum expectation values in Section \ref{secVEV} and
further need to be substituted as $\vec P\to\frac{i}{\sqrt2}\vec P$.

It is clear that all charges in ${{\cal} Q^{s-1}}$ can be expressed as
linear combinations of basis elements
\be
 I_{s-1}^{[j]}=\oint dz P_s^{[j]}\,,\,\, j=1,\ldots,\mathrm{dim{\cal Q}^{s-1}},
\ee
where $P_s^{[j]}$ is an appropriately chosen basis of monomial bosonic
currents, which span ${\cal B}^s$ modulo total derivatives. For example,
let us consider $s=4$, and for simplicity assume $N=1$ (we have only one
bosonic field). Then ${\cal B}^4$ can be spanned by
\be
 \pp^4X,\,\pp^3X\pp X,\,(\pp^2X^2),\,\pp^2X(\pp X)^2,\,(\pp X)^4.
\ee
 Taking into account that total derivatives do not contribute to the
 charges
\beq
\pp^4X&\doteq&0\nonumber\\
\pp^3X\pp X+(\pp^2X)^2&\doteq&0\nonumber\\
3\pp^2X(\pp X)^2&\doteq&0,
\eeq
we can perform a Gaussian elimination to obtain a basis modulo total
derivatives:
\beq
 P_4^{[1]}=(\pp^2X)^2.
\eeq
The above is a fairly simple example, but for larger $N$ and $s$ the
basis tends to grow rapidly, making the problem of finding commuting
charges difficult.
\begin{center}
\begin{tabular}{|c||c|c|c|c|c|c|}
\hline
s&1&3&5&7&9&11\\
\hline
$\mathrm{dim}({\cal B}^{s+1})$&1&3&7&16&35&77\\
\hline
$\mathrm{dim}({\cal Q}^{s})$&1&2&4&8&15&31\\
\hline
\end{tabular}\\
Table 1. Dimensions of the space of $O(N)$-symmetric bosonic currents
for several 
spins.
\end{center}

  If we want the currents to have some particular symmetry, say $O(N)$ or
  ${\mathbb Z_2}^{N}$,
the growth of the
basis modulo
total derivatives will be significantly reduced.

We can propose the following procedure for calculating of two
integrals of motion $I_{s_1}$ and $I_{s_2}$ of spins $s_1$ and $s_2$ with
symmetry group $G$(we will denote $s_3=s_1+s_2$):
\begin{enumerate}
\item\label{st1}  For $f\in\{1,2,3\}$ , enumerate all possible
  currents $P_{s_f}^{[j]}\in {\cal B}^{s_f}$, and
  $P_{s_f+1}^{[k]}\in{\cal B}^{s_s+1}$, compatible with $G$, where
  $j\in\{1\ldots m\equiv\dim{\cal B}^{s_f}\}$ and $k\in\{1\ldots n\equiv
   \dim{\cal
  B}^{s_f+1}\}$. Then, by ``equating'' the total derivatives $\pp
  P_{s_f}^{[j]}\doteq0$ we obtain a homogeneous linear system of $m$
  equations for $n$ variables. As we saw in the above example, by
  Gaussian eliminaion one can choose a $n-m$-dimensional basis(in fact,
  we must write $\mathrm{rank}(\pp
  I_{s_f}^{[j]}\doteq0)$, but it turns out that in the cases studied so
  far the equations were independent) for ${\cal
  B}^{s_f+1}$.
\item The integrals of motion are constructed as
\beq
\mI_{s_1}&=&\oint dz\sum_{l_1=1}^{\dim{\cal Q}^{s_1}}
 a_{l_1}P_{s_1+1}^{[l_1]}(z)\nonumber\\
\mI_{s_2}&=&\oint dw\sum_{l_2=1}^{\dim{\cal
	Q}^{s_2}}b_{l_2}P_{s_2+1}^{[l_2]}(w) 
\eeq
For each $(l_1,l_2)\in\{1\ldots\dim{\cal Q}^{s_1}\}
 \times\{1\ldots\dim{\cal Q}^{s_2}\}$ calculate the {\it residue} of the
 operator product expansion (OPE) of
 $P_{s_1+1}^{[l_1]}(z)P_{s_2+1}^{[l_2]}(w)$.
\item\label{st3} The result of the previous operation will be in ${\cal
  B}^{s_3+1}$. To obtain its value in ${\cal Q}^{s_3}$, the
  factorization is done by substituting the solution of the linear
  system obtained for $s_3$ in step \ref{st1}.
\item In order $[\mI_{s_1},\mI_{s_2}]=0$, the result from step \ref{st3}
  must be zero. By requiring that the coefficient before each
$P_{s_3+1}^{[l_3]}$ to be zero, we obtain a homogeneous bilinear
  system for
  $a_{l_1}$ and $b_{l_2}$. Although that system usually seems to be
  overdetermined(for non-trivial values of $s_1$ and $s_2$), it turned
  out that the equations are not independent and solutions were
  obtained, which will be presented in \ref{ressection}.
\end{enumerate}

Here I will address the question of implementing the above
calculations in practice. Like it was already mentioned above, the size
of the basis in ${\cal Q}^s$ grows rapidly with $s$, thus making the
systems of equations for the coefficients of commuting charges
prohibitively large not only to solve, but even to write down by
hand. That's why in order to partially cope with this complexity, it is
a good idea to develop a computer program,
which will handle
the ingenuous, though plentiful operations related to implementing the
consequence of steps described in section \ref{mathbackground}, finally
providing on its output the bilinear system of equations mentioned
afore. Further, one needs to solve that essentially quadratic algebraic
system.

For a particular symmetry, the program will have a ``combinatorial''
module, which enumerates all bosonic currents of spin $s$ with that
symmetry. We need to invoke the module also for spin $s+1$. Thus, symbol
tables are procuced for spins $s$ and $s+1$ of sizes $m$ and $n$
respectively. We need to have simple 
function which calculates the derivative of a spin-$s$ current and
expresses it as a linear combination of $s+1$ currents. Thus we have a
linear homogeneous integer-coefficient system of $m$ equations for $n$
variables. The solution of the system by Gaussian elimination and the
further calculations involve rational numbers, that's why in this
project I used the GNU Multiple Precision Arithmetic Library, {\texttt
  {gmp}}. Maybe the most crucial part of the program is the module which
actually computes the residue of the OPE of given two
currents. Essentially, one needs to count the contribution of every
possible Wick contraction. This is done with standard recursion
techniques. After that part is built, one needs to put all that
inventory together. This software\footnote{The
source code is available at
\href{http://www.physics.rutgers.edu/~vitchev/ope/}
{http://www.physics.rutgers.edu/$\sim$vitchev/ope/}}
 is enough flexible, so that by 
providing a different ``combinatorial'' module, and with some
modifications to the OPE module, one might study cases of different
symmetry groups or 
related problems involving operator product expansions of currents.

\section{Several Families of Commuting Charges}\label{ressection}
\subsection{$O(N)$ symmetry}
We consider $N$ free bosonic fields $X^\mu$, $\mu\in\{1\ldots N\}$. The
currents can be constructed as polynomials in terms of the form
$\pp^{n_1}X\cdot\pp^{n_2}X$. It is convenient to use the notation
\be
(n_1,n_2)\equiv\pp^{n_1}X\cdot\pp^{n_2}X
 =\sum_{\mu=1}^N\pp^{n_1}X^\m
  \pp^{n_2}X^\n.
\ee
Where there is no possibility for ambiguity, one can write simply
$(n_1n_2)$.

A suitable basis for ${\cal Q}^3$ is
\beq
\I^{[0]}_4&=&(22)\nonumber\\
\I^{[1]}_4&=&(11)(11).
\eeq

Correspondingly, possible bases for ${\cal Q}^5$, ${\cal Q}^7$ and
${\cal Q}^9$ are
\be
\begin{tabular}{|lcr|lcr|}
\hline
$\I^{[0]}_6$&$=$&$(33)$&$\I^{[2]}_6$&$=$&$(12)(12)$\\
$\I^{[1]}_6$&$=$&$(22)(11)$&$\I^{[3]}_6$&$=$&$(11)(11)(11)$\\
\hline
\end{tabular}
\ee
\be
\begin{tabular}{|rcl|rcl|}
\hline
$\I^{[0]}_8$&$=$&$(23)(12)$&$\I^{[4]}_8$&$=$&$(44)$\\
$\I^{[1]}_8$&$=$&$(13)(13)$&$I^{[5]}_8$&$=$&$(22)(11)(11)$\\
$\I^{[2]}_8$&$=$&$(33)(11)$&$\I^{[6]}_8$&$=$&$(12)(12)(11)$\\
$\I^{[3]}_8$&$=$&$(22)(22)$&$\I^{[7]}_8$&$=$&$(11)(11)(11)(11)$\\
\hline
\end{tabular}
\ee
\be
\begin{tabular}{|lcr|lcr|}
\hline
$\I^{[0]}_{10}$&$=$&$(23)(23)$&$\I^{[8]}_{10}$&$=$&$(22)(22)(11)$\\
$\I^{[1]}_{10}$&$=$&$(55)$&$\I^{[9]}_{10}$&$=$&$(23)(14)$\\
$\I^{[2]}_{10}$&$=$&$(44)(11)$&$\I^{[10]}_{10}$&$=$&$(22)(12)(12)$\\
$\I^{[3]}_{10}$&$=$&$(33)(11)(11)$&$\I^{[11]}_{10}$&$=$&$(14)(14)$\\
$\I^{[4]}_{10}$&$=$&$(24)(13)$&$\I^{[12]}_{10}$&$=$&$(22)(11)(11)(11)$\\
$\I^{[5]}_{10}$&$=$&$(23)(12)(11)$&$\I^{[13]}_{10}$&$=$&$(12)(12)(11)(11)$\\
$\I^{[6]}_{10}$&$=$&$(13)(13)(11)$&$\I^{[14]}_{10}$&$=$&$(11)(11)(11)(11)(11)$\\
$\I^{[7]}_{10}$&$=$&$(33)(22)$&\ &\ &\ \\
\hline
\end{tabular}
\ee

There are several interesting families of commuting charges. Here we
will give the results for the several first members of each family:
\paragraph{(1)KdV-Virasoro} This series\footnote{The first member in
  this series as well as in the other series in this paper is $\mI_1=\oint
  dz\,T(z)$, where $T$ is the energy-momentum tensor.} is related to boundary 
Sine-Gordon model \cite{Sasaki:1987mm}
\cite{Eguchi:1989hs} \cite{Bazhanov:1994ft},
 and can
also be written in terms of 
composite fields built of the 
energy-momentum tensor-- see \eqref{bCMP}-\eqref{eCMP}. 
\be
\mI_3=\oint dz\,-2\I^{[0]}_4+\I^{[1]}_4
\ee
\be
\mI_5=\oint dz\,\frac1{18}(56+N)\I^{[0]}_6
 -6\I^{[1]}_6
 -\frac23(20+N)\I^{[2]}_6
 +\I^{[3]}_6
\ee
\beq
\mI_7&=&\oint dz\,-\frac8{45}(2+N)(39+2N)\I^{[0]}_8
 +\frac4{45}(364+34N+N^2)\I^{[1]}_8\nonumber\\
 &&+\frac29(62+N)\I^{[2]}_8
 -\frac4{45}(229+34N+N^2)\I^{[3]}_8
 -\frac2{657}(1464+59N+N^2)\I^{[4]}_8\nonumber\\
 &&-12\I^{[5]}_8
 -\frac83(26+N)\I^{[6]}_8
 +\I^{[7]}_8
\eeq
\paragraph{(2)''$O(N)$''} This is the family of commuting integrals of
motion associated with spherical branes found recently in 
\cite{Lukyanov:2003rt}. It 
cannot be expressed completely in terms of $T$:
\be
\mI_3=\oint dz\,-\frac23(2+N)\I^{[0]}_4+\I^{[1]}_4
\ee
\be
\mI_5=\oint dz\,\frac1{150}(4+N)(59+36N)\I^{[0]}_6
 -\frac65(4+N)\I^{[1]}_6
 -\frac{14}5(4+N)\I^{[2]}_6
 +\I^{[3]}_6
\ee
\beq
\mI_7&=&\oint dz\,-\frac8{735}(6+N)(199+88N)\I^{[0]}_8
 +\frac4{735}(6+N)(692+239N)\I^{[1]}_8\nonumber\\
 &&+\frac2{49}(6+N)(37+12N)\I^{[2]}_8
 -\frac8{735}(6+N)(211+97N)\I^{[3]}_8\nonumber\\
 &&-\frac2{77175}(6+N)(11009+11183N+2700N^2)\I^{[4]}_8
 -\frac{12}7(6+N)\I^{[5]}_8\nonumber\\
 &&-\frac{72}7(6+N)\I^{[6]}_8
 +\I^{[7]}_8
\eeq
\paragraph{(3)''$\Phi_{1,2}$'' or $A_2^{(2)}$}
This family has spins $1,6n\pm1$, and is related to a reduction of the
quantum Boussinesq theory \cite{Bazhanov:2001xm} and boundary
Bullough-Dodd model 
\cite{Dodd:bi} \cite{Zhiber:am}:
\beq
\mI_5&=&\oint dz\,\frac1{96}\(107+17N+15\sqrt{(N-25)(N-1)}\)\I^{[0]}_6
 -6\I^{[1]}_6\nonumber\\
 &&-\frac18\(-85+17N+15\sqrt{(N-25)(N-1)}\)\I^{[2]}_6
 +\I^{[3]}_6
\eeq
\beq
\mI_7&=&\oint dz\,\nonumber\\
 &&\frac1{120}
  \(3621+4160N-325N^2+(609+315N)\sqrt{(N-25)(N-1)}\)\I^{[0]}_8\nonumber\\
 &&\frac1{480}
  \(-3171-2810N+325N^2+(441+315N)\sqrt{(N-25)(N-1)}\)\I^{[1]}_8\nonumber\\
 &&+\frac1{16}\(131+9N+7\sqrt{(N-25)(N-1)}\)\I^{[2]}_8\nonumber\\
 &&+\frac1{480}\(8931+2810N-325N^2+(315N-441)
  \sqrt{(N-25)(N-1)}\)\I^{[3]}_8\nonumber\\
 &&+\frac1{14400}\(-16929+110N-325N^2+(315N-2541)\sqrt{(N-25)(N-1)}\)
  \I^{[4]}_8\nonumber\\
 &&-12\I^{[5]}_8\nonumber\\
 &&-\frac34\(3+9N+7\sqrt{(N-25)(N-1)}\)\I^{[6]}_8\nonumber\\
 &&+\I^{[7]}_8
\eeq
The densities in this series can be written as
\beq
P_6&=&8T^3+\frac{181-17N-15\sqN}8(\pp T)^2
\eeq
\beq
P_8&=&16T^4+\frac34\(61-9N-7\sqN\)2T(\pp T)^2\nonumber\\
 &+&\frac{325N^2-6590N+12849+(315N-2499)\sqN}{480}T\pp^4 T,
\eeq
where we use the regular part of the composite fields built from
derivatives of the energy-momentum tensor:
(cf. \eqref{compFields})
\be\label{bCMP}
T=\frac12(11)
\ee
\be
\pp T=(12)
\ee
\be
2T^2=(11)(11)-2(22)
\ee
\be
8T^3=(11)(11)(11)-6(22)(11)-12(12)(12)+3(33)
\ee
\be
(\pp T)^2=(12)(12)-\frac1{12}(33)
\ee
\vskip 0.5cm
\beq
16T^4&=&(11)^4-12(22)(11)(11)-48(12)(12)(11)+12(33)(11)\nonumber\\
 &&+20(13)(13)-8(22)(22)
  -\frac{10}3(44)
\eeq
\be
2T(\pp T)^2=(12)(12)(11)+3(23)(12)-\frac76(13)(13)+\frac76(22)(22)
  -\frac1{12}(33)(11)+\frac1{15}(44)
\ee
\be\label{eCMP}
2T\pp^4T=2(13)(13)-\frac1{15}(44)-8(23)(12)-2(22)(22)
\ee

\subsection{$\mathbb Z_2\times\mathbb Z_2 $ symmetry}\label{secZ2Z2}
Here we will use the notation
\be
(n_1,n_2,\ldots,n_\alpha|m_1,m_2,\ldots,m_\beta)\equiv
 \pp^{n_1}X\pp^{n_2}X\ldots\pp^{n_\alpha}X
  \pp^{m_1}Y\pp^{m_2}Y\ldots\pp^{m_\beta}Y.
\ee
Where no ambiguity is possible, we will omit the commas.
Possible choice of bases for two bosonic fields with $\mathbb
Z_2\times\mathbb Z_2 $ symmetry is
\be
\begin{tabular}{|lcr|lcr|}
\hline
$\I^{[0]}_4$&$=$&$(|22)$&$\I^{[3]}_4$&$=$&$(11|11)$\\
$\I^{[1]}_4$&$=$&$(22|)$&$\I^{[4]}_4$&$=$&$(1111|)$\\
$\I^{[2]}_4$&$=$&$(|1111)$&\ &\ &\\
\hline
\end{tabular}
\ee

\be
\begin{tabular}{|lcr|lcr|}
\hline
$\I^{[0]}_6$&$=$&$(33|)$&$\I^{[6]}_6$&$=$&$(1122|)$\\
$\I^{[1]}_6$&$=$&$(11|22)$&$\I^{[7]}_6$&$=$&$(|111111)$\\
$\I^{[2]}_6$&$=$&$(12|12)$&$\I^{[8]}_6$&$=$&$(11|1111)$\\
$\I^{[3]}_6$&$=$&$(|33)$&$\I^{[9]}_6$&$=$&$(1111|11)$\\
$\I^{[4]}_6$&$=$&$(22|11)$&$\I^{[10]}_6$&$=$&$(111111|)$\\
$\I^{[5]}_6$&$=$&$(|1122)$&\ &\ &\\
\hline
\end{tabular}
\ee

\be
\begin{tabular}{|rcl|rcl|}
\hline
$\I^{[0]}_8$&$=$&$(33|11)$&$\I^{[13]}_8$&$=$&$(22|22)$\\
$\I^{[1]}_8$&$=$&$(44|)$&$\I^{[14]}_8$&$=$&$(1111|22)$\\
$\I^{[2]}_8$&$=$&$(13|13)$&$\I^{[15]}_8$&$=$&$(1112|12)$\\
$\I^{[3]}_8$&$=$&$(1133|)$&$\I^{[16]}_8$&$=$&$(23|12)$\\
$\I^{[4]}_8$&$=$&$(1223|)$&$\I^{[17]}_8$&$=$&$(1122|11)$\\
$\I^{[5]}_8$&$=$&$(|44)$&$\I^{[18]}_8$&$=$&$(|1223)$\\
$\I^{[6]}_8$&$=$&$(14|12)$&$\I^{[19]}_8$&$=$&$(111122|)$\\
$\I^{[7]}_8$&$=$&$(|111122)$&$\I^{[20]}_8$&$=$&$(|11111111)$\\
$\I^{[8]}_8$&$=$&$(|1133)$&$\I^{[21]}_8$&$=$&$(11|111111)$\\
$\I^{[9]}_8$&$=$&$(11|1122)$&$\I^{[22]}_8$&$=$&$(1111|1111)$\\
$\I^{[10]}_8$&$=$&$(12|1112)$&$\I^{[23]}_8$&$=$&$(111111|11)$\\
$\I^{[11]}_8$&$=$&$(11|33)$&$\I^{[24]}_8$&$=$&$(11111111|)$\\
$\I^{[12]}_8$&$=$&$(22|1111)$&$$&$$&$$\\
\hline
\end{tabular}
\ee

It turns out that the ``$O(N)$''(when $N=2$) family has a
$\mathbb Z_2\times\mathbb Z_2 $-symmetric deformation, related to the
``paperclip'' model \cite{Lukyanov:2003nj}. Here are the densities of some of
the commuting integrals of motion of that
family:

\beq
P_4&=&-\frac{4n^2+7n+2}{3(3n+2)(3n+4)}\I_6^{[0]}
 -\frac{4n^2+9n+4}{3(3n+2)(3n+4)}\I_6^{[1]}\nonumber\\
&&+\frac{n+2}{6(3n+4)}\I_6^{[2]}
 +\frac{n(n+2)}{(3n+2)(3n+4)}\I_6^{[3]}
 +\frac{n}{6(3n+2)}\I_6^{[4]}
\nonumber\\
\eeq
\beq
P_6&=&-\frac{96+500n+860n^2+575n^3+131n^4}{120n(2+n)}
    \I^{[0]}_6\nonumber\\
 &&+\frac{2+11n+6n^2}{4}\I^{[1]}_6\nonumber\\
 &&+(4+14n+7n^2)\I^{[2]}_6\nonumber\\
 &&-\frac{32+232n+554n^2+473n^3+131n^4}{120n(2+n)}
    \I^{[3]}_6\nonumber\\
 &&+\frac{4+13n+6n^2}{4}\I^{[4]}_6\nonumber\\
 &&+\frac{(2+5n)(4+9n+4n^2)}{4n}\I^{[5]}_6\nonumber\\
 &&+\frac{(8+5n)(2+7n+4n^2)}{4(2+n)}\I^{[6]}_6\nonumber\\
 &&-\frac{(2+n)(2+5n)(4+5n)}{120n}\I^{[7]}_6\nonumber\\
 &&-\frac{(2+n)(2+5n)}{8}\I^{[8]}_6\nonumber\\
 &&-\frac{n(8+5n)}{8}\I^{[9]}_6\nonumber\\
 &&-\frac{(6+5n)(8+5n)}{120(2+n)}\I^{[10]}_6
\eeq

\subsection {The $\mathbb Z_2\times O(N)$ (``cylindrical'') case}
Let us consider $N+1$ bosonic fields $(X,Y_1,\ldots Y_N)\equiv(X,\vec
Y)$. There are three 
non-trivial solutions for $I_3$ and $I_5$ with $\mathbb Z_2\times O(N)$,
and it seems plausible that these are the first members of infinite
families of commuting charges. The results can be expressed in terms of
only $X$ and the 
regular part of the composite fields of the energy-momentum tensor of
$\vec Y$. Namely, these composite fields are
\beq\label{compFields}
T_Y&=&\frac12\pp \vec Y\cdot\pp\vec Y\nonumber\\
\pp T_Y&=&\pp\vec Y\cdot\pp^2\vec Y\nonumber\\
2T_Y^2&=&(\pp \vec Y\cdot\pp\vec Y)^2-2\pp^2\vec Y\cdot\pp^2\vec
 Y\nonumber\\
8T_Y^3&=&(\pp\vec Y\cdot\pp\vec Y)^3-6(\pp^2\vec Y\cdot\pp^2\vec Y)
  (\pp\vec Y\cdot\pp\vec Y)
  -12(\pp\vec Y\cdot\pp^2\vec Y)^2+3\pp^3\vec Y\cdot\pp^3\vec Y\nonumber\\
(\pp T)^2&=&(\pp\vec Y\cdot\pp^2\vec Y)^2-\frac1{12}\pp^3\vec
  Y\cdot\pp^3\vec Y.
\eeq
In terms of those composite fields, the 
currents can be written as:

Solution 1:
One solution (Solution 1) is
\beq
P_4&=&4T_Y^2+\frac{1}{6}\(-9-3N-\sqN\)(\pp X)^2\nonumber\\
&& +12(\pp X)^2T_Y+\frac{1}3\(2+N+\sqN\)(\pp X)^4
\eeq
\beq
P_6&=&\frac8{15}T_Y^3-\frac{34+2N}{45}(\pp T_Y)^2+4(\pp
 X)^2T_Y^2\nonumber\\
&&+\frac{\(307+106N+19N^2+(53+13N)\sqN\)}{1620}(\pp^3X)^3\nonumber\\
&& -\frac{4\(16+2N+\sqN\)}{9}\pp X\pp^2X\pp T_Y\nonumber\\
&&-\frac{\(33+3N+\sqN\)}{9}(\pp^2X)^2T_Y\nonumber\\
&&-\frac{\(131-31N+8N^2+(31+8N)\sqN\)}{54}(\pp
 X)^2(\pp^2X)^2\nonumber\\
&&\frac{2\(8+N+\sqN\)}{9}(\pp X)^4T_Y\nonumber\\
&&\frac{\(58-35N+4N^2+(17+4N)\sqN\)}{405}(\pp X)^6.
\eeq
When $N=1$, this solution coincides with the paperclip series with
$n=-1$.

A second solution (Solution 2) is
\beq
P_4&=&4T_Y^2-\frac{\(409+46N+N^2+(-5+N)\sqN\)}{192}(\pp^2
 X)^2\nonumber\\
&&+\frac{\(11+N+\sqN\)}{4}(\pp X)^2T_Y\nonumber\\
&&+\frac{\(117+2N+N^2+(15+N)\sqN\)}{192}(\pp X)^4
\eeq

\beq
P_6&=&\frac{8\(20+N-2\sqN\)}{15(2+N)}T_Y^3+\frac{\(11+N+\sqN\)}{8}(\pp X)^4T\nonumber\\
&&+\frac{-125-7N+9\sqN}{120}(\pp T_Y)^2
+4(\pp X)^2T_Y^2\nonumber\\
&&+\frac{6930+1265N+218N^2+11N^3+(778+111N+5N^2)\sqN}{1440(2+N)}
  \pp^3 X^2\nonumber\\
&&-\frac{2(14+N)}{3}\pp X\pp^2X\pp T_Y
 -\frac{\(101+7N-\sqN\)}{24}(\pp^2 X)^2T_Y\nonumber\\
&&-\frac{\(505+68N+3N^2+(27+3N)\sqN\)}{96}(\pp X)^2(\pp^2X)^2\nonumber\\
&&\frac{b_9\(95+N^2+(13+N)\sqN\)}{480}(\pp X)^6.
\eeq
For $N=1$, this solution coincides with the paperclip solution, but with $n=2$.

There is also a third solution (Solution 3), namely
\beq
P_4&=&4T_Y^2+\frac{\(13-27N+2N^2+(5-2N)\sqN\)}{6}(\pp^2
 X)^2\nonumber\\
&&+2\(7-N+\sqN\)(\pp X)^2T_Y+(\pp X)^4
\eeq
\beq
P_6&=&-\frac{8\(-5+2N+\sqN\)}{15(2+N)}T_Y^3\nonumber\\
&&+\frac{-15+3N-\sqN}{15}(\pp T_Y)^2+4(\pp X)^2T_Y^2\nonumber\\
&&+\frac{190-61N-7N^2+22N^3+(-38-41N-14N^2)\sqN}{180(2+N)}\pp^3
 X^2\nonumber\\
&&-\frac{2\(9+3N-\sqN\)}{3}\pp X\pp^2X\pp T_Y\nonumber\\
&&-\frac{7+5N-\sqN}{3}(\pp^2X)^2T_Y\nonumber\\
&&-\frac{40-5N+N^2+(-8-N)\sqN}{6(2+N)}(\pp X)^2(\pp^2X)^2
 +2(\pp X)^4T_Y\nonumber\\
&&-\frac{-5+2N+\sqN}{15(2+N)}(\pp X)^6
\eeq
Like Solution 1, for $N=1$ this one coincides with the ``paperclip''
solution with $n=-1$.



Summarily, $N=1$, both Solution 1 and Solution 3 coincide with the
``paperclip'' series with $n=-1$, and Solution 2 coincides with the
``paperclip'' series as well, but with $n=2$.

\section{Vacuum expectation values}\label{secVEV}
The vaucuum expectation values of the $\mathbb Z_2\times\mathbb
Z_2$-symmetric integrals of motion from Section $\ref{secZ2Z2}$ are: 
\beq
I_3&=&\frac{4+3n}{6(2+n)}P^4+\frac{2+3n}{6n}Q^4
 +P^2Q^2+\nonumber\\
 &&\frac{3+2n}{6(2+n)}P^2+\frac{1+2n}{6n}Q^2
  +\frac{11+36n+18n^2}{360n(2+n)}
\eeq
\beq
I_5&=&-\frac{(6+5n)(8+5n)}{120(2+n)^2}P^6
  -\frac{(2+5n)(4+5n)}{120n^2}Q^6\nonumber\\
 &&-\frac{8+5n}{8(2+n)}P^4Q^2-\frac{2+5n}{8n}P^2Q^2\nonumber\\
 &&-\frac{(4+3n)(8+5n)}{48(2+n)^2}P^4-\frac{(2+3n)(2+5n)}{48n^2}Q^4
  -\frac{2+10n+5n^2}{8n(2+n)}P^2Q^2\nonumber\\
 &&-\frac{30+225n+250n^2+76n^3}{480n(2+n)^2}P^2
  -\frac{28+137n+206n^2+76n^3}{480n^2(2+n)}Q^2\nonumber\\
 &&-\frac{564+3410n+7385n^2+5680n^3+1420n^4}{60480n^2(2+n)^2}
\eeq
\beq
I_7&=&\frac{n(8+7n)(10+7n)}{42(2+n)^2(2+7n)}P^8
  +\frac{(2+n)(4+7n)(6+7n)}{42n^2(12+7n)}Q^8\nonumber\\
 &&+\frac{2n(10+7n)}{3(2+n)(2+7n)}P^6Q^2
  +\frac{2(2+n)(4+7n)}{3n(12+7n)}P^2Q^6+P^4Q^4\nonumber\\
 &&+\frac{n(5+4n)(10+7n)}{9(2+n)^2(2+7n)}P^6
  +\frac{(2+n)(3+4n)(4+7n)}{9n^2(12+7n)}Q^6\nonumber\\
 &&+\frac{6+49n+28n^2}{3(2+n)(2+7n)}P^4Q^2
  +\frac{20+63n+28n^2}{3n(12+7n)}P^2Q^4\nonumber\\
 &&+\frac{122+1407n+1792n^2+612n^3}{180(2+n)^2(2+7n)}P^4
  +\frac{420+1583n+1880n^2+612n^3}{180n^2(12+7n)}Q^4\nonumber\\
 &&+\frac{7(40+338n+985n^2+816n^3+204n^3)}
    {30n(2+n)(2+7n)(12+7n)}P^2Q^2\nonumber\\
 &&+\frac{2520+26166n+102459n^2+136612n^3+74676n^4+14552n^5}
    {1260n(2+n)^2(2+7n)(12+7n)}P^2\nonumber\\
 &&+\frac{3720+30202n+89149n^2+121284n^3+70844n^4+14552n^5}
    {1260n^2(2+n)(2+7n)(12+7n)}Q^2\nonumber\\
 &&+\frac{68760+632142n+2264647n^2+4095840n^3+3708040n^4+1610448n^5+268408n^6}
    {151200n^2(2+n)^2(2+7n)(12+7n)}\nonumber\\
\eeq

\section{The vacuum expectation values and the quasiclassical approximation}
In this section we will demonstrate the ODE/IM correspondence for the
``paperclip'' series from Section \ref{secZ2Z2}. The corresponding equation
\cite{Lukyanov:2003nj} is: 
\be\label{Leq1}
-\pp_z^2\Psi(z)+\kappa^2(1+e^z)^n\Psi(z)+\frac{nP^2}2\frac{e^z}{1+e^z}\Psi(z)
 +\(\frac{n+2}2Q^2+\frac14\)\Psi(z)=0.
\ee

One can try to calculate the Wronskian of \eqref{Leq1} in the
``quasiclassical'' limit $\kappa\to\infty$. To this end, we use the
ansatz
\be
\Psi(z)=\exp[\kappa S_{-1}(z)+S_0(z)+\frac1\kappa S_{1}(z)
 +\frac1{\kappa^2}S_{2}(z)+\ldots]
\ee

It is possible to show that the expansion of the Wronskian is
\be
W[f_+,f_-]=\exp[\kappa S_{-1}(-\infty)+S_0(z)+\frac1\kappa
  S_{1}(-\infty)
 +\frac1{\kappa^3}S_{3}(-\infty)+\frac1{\kappa^5}S_{5}(-\infty)
 +\ldots].
\ee

Calcultions yield:
\beq
S_1^{(-\infty)}&=&\frac{P^2}2+\frac{Q^2}2+\frac18
\eeq
\beq
S_3^{(-\infty)}&=&-\frac{n}{24(2+3n)}P^4-\frac{2+n}{24(4+3n)}Q^4
  -\frac{n(2+n)}{4(2+3n)}P^2Q^2\nonumber\\
 &-&\frac{n(3+2n)}{24(2+3n)(4+3n)}P^2-\frac{(2+n)(1+2n)}{24(2+3n)(4+3n)}Q^2
  -\frac{2+6n+3n^2}{192(2+3n)(4+3n)}
\eeq
\beq
S_5{(-\infty)}&=&\frac{n^2}{40(2+5n)(4+5n)}P^6
    +\frac{(2+n)^2}{40(6+5n)(8+5n)}Q^6\nonumber\\
 &+&\frac{3n^2(2+n)}{8(2+5n)(4+5n)(6+5n)}P^4Q^2
  +\frac{3n(2+n)^2}{8(4+5n)(6+5n)(8+5n)}P^2Q^4\nonumber\\
 &+&\frac{n^2(4+3n)}{16(2+5n)(4+5n)(6+5n)}P^4
  +\frac{(2+n)^2(2+3n)}{16(4+5n)(6+5n)(8+5n)}Q^4\nonumber\\
 &+&\frac{3n(2+n)(2+10n+5n^2)}{8(2+5n)(4+5n)(6+5n)(8+5n)}P^2Q^2\nonumber\\
 &+&\frac{n(30+225n+250n^2+76n^3)}{160(2+5n)(4+5n)(6+5n)(8+5n)}P^2\nonumber\\
 &+&\frac{(2+n)(28+137n+206n^2+76n^3)}
  {160(2+5n)(4+5n)(6+5n)(8+5n)}Q^2\nonumber\\
 &+&\frac{24+140n+290n^2+220n^3+55n^4}{640(2+5n)(4+5n)(6+5n)(8+5n)}
\eeq
\beq
S_7&=&-\frac{15n^3}{448(2+7n)(4+7n)(6+7n)}P^8
  -\frac{15(2+n)^3}{448(8+7n)(10+7n)(12+7n)}Q^8\nonumber\\
 &-&\frac{15n^3(2+n)}{16(2+7n)(4+7n)(6+7n)(8+7n)}P^6Q^2\nonumber\\
 &-&\frac{15n(2+n)^2}{16(6+7n)(8+7n)(10+7n)(12+7n)}P^2Q^6\nonumber\\
 &-&\frac{45n^2(2+n)^2}{32(4+7n)(6+7n)(8+7n)(10+7n)}P^4Q^4\nonumber\\
 &-&\frac{5n^3(5+4n)}{32(2+7n)(4+7n)(6+7n)(8+7n)}P^6\nonumber\\
 &-&\frac{5(2+n)^3(3+4n)}{32(6+7n)(8+7n)(10+7n)(12+7n)}Q^6\nonumber\\
 &-&\frac{15n^2(2+n)(6+49n+28n^2)}
    {32(2+7n)(4+7n)(6+7n)(8+7n)(10+7n)}P^4Q^2\nonumber\\
 &-&\frac{15n(2+n)^2(20+63n+28n^2)}
    {32(4+7n)(6+7n)(8+7n)(10+7n)(12+7n)}P^2Q^4\nonumber\\
 &-&\frac{n^2(122+1407n+1792n^2+612n^3)}
    {128(2+7n)(4+7n)(6+7n)(8+7n)(10+7n)}P^4\nonumber\\
 &-&\frac{(2+n)^2(420+1583n+1880n^2+612n^3)}
    {128(4+7n)(6+7n)(8+7n)(10+7n)(12+7n)}Q^4\nonumber\\
 &-&\frac{21n(2+n)(40+338n+985n^2+816n^3+204n^4)}
    {64(2+7n)(4+7n)(6+7n)(8+7n)(10+7n)(12+7n)}P^2Q^2\nonumber\\
 &-&\frac{n(2520+26166n+102459n^2+136612n^3+74676n^4+14552n^5)}
    {896(2+7n)(4+7n)(6+7n)(8+7n)(10+7n)(12+7n)}P^2\nonumber\\
 &-&\frac{(2+n)(3720+30202n+89149n^2+121284n^3+70844n^4+14552n^5)}
    {896(2+7n)(4+7n)(6+7n)(8+7n)(10+7n)(12+7n)}Q^2\nonumber\\
 &-&\frac{12240+110628n+386834n^2+680568n^3+606452n^4+261786n^5+43631n^6}
    {14336(2+7n)(4+7n)(6+7n)(8+7n)(10+7n)(12+7n)}
\eeq
One can observe that
\beq
  S_1&=&\alpha_1 I_1+\beta_1\nonumber\\
  S_3&=&\alpha_3 I_3+\beta_3\nonumber\\
  S_5&=&\alpha_5 I_5+\beta_5\nonumber\\
  S_7&=&\alpha_7 I_7+\beta_7\nonumber\\
\eeq
where $\alpha_s$ are normalization factors, and 
\beq
 \beta_1&=&\frac1{24}\nonumber\\
 \beta_3&=&-\frac1{2880}\nonumber\\
 \beta_5&=&\frac1{40320}\nonumber\\
 \beta_7&=&-\frac1{215040}
\eeq
turn out to be coefficients of the expansion
\be
\log\Gamma(2\kappa)=\ldots+\sum_{k=1}^\infty\frac{B_{2k}}
    {2k(2k-1)2^{2k-1}\kappa^{2k-1}},
\ee
\be
\b_{2k-1}=\frac{B_{2k}}{2k(2k-1)2^{2k-1}},
\ee
where $B_{2k}$ are the Bernoulli numbers.

Thus, for the Wronskian we get
\be
W[f_+,f_-]\sim\Gamma[2\kappa]\exp\[\frac{ \tilde I_1}\kappa
+\frac{ \tilde I_3}{\kappa^3}
+\frac{ \tilde I_5}{\kappa^5}
+\ldots
\]
\ee

\section{Conclusions}
Commuting integrals of motion are an important feature of integrable
field theories, and, in particular, conformal field theories which
allow integrable boundary interaction. In that case, the associated
boundary state in such a theory preserves the corresponding family of
bulk IM. Moreover, the boundary state, namely its overlap amplitudes
with states of the Fock space of the bulk theory, can be expressed in
terms of eigenvalues of the integrals of motion 
\cite{Bazhanov:2003ni,Bazhanov:1994ft, Bazhanov:1996dr}. In order to
find families of commuting integrals of motion, associated with such
theories, I designed several computer programs (available on demand), which are
performing the necessary calculation of commutators of homogeneous
bosonic integrals 
of motion. 
In this work we showed the first members of several families of
commuting bosonic integrals 
of motion, obtained as solutions of the equations generated by the
programs in question, with
symmetries $O(N)$, $\mathbb Z_2\times\mathbb Z_2$, and 
$\mathbb Z_2\times O(N)$. Some of these families are identified as being
associated with a particular boundary interaction-- the $\mathbb
Z_2\times \mathbb Z_2$ one with the ``paperclip'' \cite{Lukyanov:2003nj}
and the  
$O(N)$ family is related to ``$O(N)$ spherical brane model''
\cite{Lukyanov:2003rt}. Also, the KdV-Virasoro family is related to
boundary Sine-Gordon model \cite{Sasaki:1987mm} \cite{Eguchi:1989hs}
\cite{Bazhanov:1994ft},
and the Boussinesq family is
related to boundary Bullough-Dodd model 
\cite{Dodd:bi} \cite{Zhiber:am}.
For some other families, namely the three families with 
$\mathbb Z_2\times O(N)$ symmetry, it is not yet very clear what is the
boundary interaction associated with them.

Possible future developments might include non-conformal, but still
integrable bulk theories, like the sausage model
\cite{Fateev:1992tk}, or supersymmetric generalization.

\section*{Acknowledgements}

\noindent I would like to express my profound
gratitude to A.~B.~Zamolodchikov for his precious guidance, inspiration
and enlightening discussions. I am also grateful to S.~L.~Lukyanov for
the helpful discussions.

\end{document}